\documentclass[12pt]{article}
\usepackage{graphicx}
\usepackage{subfigure}
\begin{document}
\begin{titlepage}

\begin{center}
	{\Large {\bf A New Measurement of Dynamic Critical exponent of Wolff Algorithm by Dynamic Finite Size Scaling }}
\end{center}

\vskip 1.5cm

\centerline{
	MEHMET D\.ILAVER$^\dagger$, SEMRA G\"UND\"U\c{C}$^\dagger$, MERAL AYDIN and     
	Y\.I\u{G}\.IT G\"UND\"U\c{C}$^\dagger$
}

\centerline{\it$^\dagger$ Hacettepe University, Physics Department,}
\centerline{\it 06800  Beytepe, Ankara, Turkey }

\vskip 0.5cm

\centerline{\normalsize {\bf Abstract} }

{\small

In this work we have calculated the dynamic critical exponent $z$ for
$2$-, $3$- and $4$-dimensional Ising models using the Wolff's
algorithm through dynamic finite size scaling. We have studied time
evolution of the average cluster size, the magnetization and higher
moments of the magnetization. It is observed that dynamic scaling is
independent of the algorithm. In this sense, universality is established
for a wide range of algorithms with their own dynamic critical exponents.
For scaling, we have used the literature values of critical exponents
to observe the dynamic finite size scaling and to obtain the value of
$z$.  From the simulation data a very good scaling is observed leading
to vanishingly small $z$ values for all three dimensions.}

\vskip 0.5cm

{PACS numbers : 05.50.+q,05.10.Ln}

\end{titlepage}        

\pagebreak

\section{Introduction}

One approach to observe the critical behavior of the systems
exhibiting second-order phase transitions is to use dynamic scaling
which exists in the early stages of the quenching process in the
system. Jansen, Schaub and Schmittmann~\cite{Janss89} showed
that for a dynamic relaxation process, in which a system is evolving
according to the dynamics of Model A~\cite{Hoh77} and is quenched from a
very high temperature to the critical temperature, a universal dynamic
scaling behavior within the short-time regime
exists~\cite{Wang97,Zheng98,Zheng00}.

$\;$

For the $k^{th}$ moment of the magnetization of a system, the dynamic
finite size relation can be written as~\cite{Janss89}
\begin{equation}
	\label{Magnetization}
	M^{(k)}(t,\epsilon,{m_0,L})=L^{(-k{\beta}/{\nu})}
	{\cal{M}}^{(k)}(t/\tau,{\epsilon}L^{1/{\nu}},{m_{0}}L^{x_0})
\end{equation}
where $L$ is the spatial size of the system, $\beta$ and $\nu$ are the
well-known critical exponents, ${\epsilon}=(T-{T_c})/{T_c}$ is the
reduced temperature, $x_0$ is a new and independent exponent which is
the anomalous dimension of the initial magnetization $m_0$, $t$ is the
simulation time and $\tau$ is the autocorrelation time. For second-order 
phase transitions autocorrelation time is given as $\tau \sim
L^z$, where $z$ is the dynamic critical exponent.

$\;$

The dynamic finite size scaling requires scaling of the Monte Carlo
time as well as the thermodynamic quantities. In many different
dynamic finite size scaling studies Eq.(\ref{Magnetization}) has
been used in order to calculate the dynamic critical exponent as well as
the known critical exponents. In these studies it has been shown that
for a given local algorithm, dynamic critical exponents ($z$) are the
same within the error limits for both early stages of the simulation
and after the thermal equilibrium is
reached~\cite{Zheng98,Zheng00,Jaster99,Luo97,Ying2001,Dilaver2003}.  
Since using an algorithm
with a well-established dynamics and dynamic critical exponent
simplifies the scaling in all studies on the dynamic
finite size scaling, the simulation algorithm is chosen as one
of the local algorithms. In these studies it is
established that the dynamic finite size scaling is independent of the
algorithm, in this sense a universality is shown.  Similarly,
dynamic finite size scaling is tested on first-order phase transitions
where the critical behavior is governed by the dimension rather than
an exponent.  In the case of first-order phase transition, the
expected dynamic finite size scaling is also 
observed~\cite{Ozoguz2000,Schuelke2000}.

$\;$

Introduction of cluster algorithms~\cite{SwendsenWang,Wolff} made
great improvement in the simulations of magnetic spin
systems. Despite the fact that Wolff's algorithm is a modification of
the Swendsen-Wang algorithm, Wolff's algorithm~\cite{Wolff} exhibits
an important difference compared to other algorithms~\cite{Ito1990}.
In local algorithms and the Swendsen-Wang
algorithm~\cite{SwendsenWang}, at each Monte Carlo sweep there is an attempt
to update all the spins on the lattice. On the contrary, in
Wolff's algorithm only spins belonging to a certain cluster around the
seed spin are considered and updated at each Monte Carlo sweep. In
equilibrium simulation studies of the systems exhibiting second-order
phase transitions, the dynamic critical exponent of the
Wolff's algorithm can not be obtained directly from the measured
autocorrelation times (${\tau}^{'}$) since only a fraction of the
spins is updated. The efficiency of the Wolff's algorithm is directly
related to the average cluster size ($<C>$) at a given
temperature. Hence, for comparison of the efficiencies between
different algorithms and of the Wolff's algorithm, the average cluster
size plays an important role. For Wolff's algorithm the ratio between
the observed autocorrelation time and the actual autocorrelation time
is proportional to the average cluster size.

$\;$

In the literature the dynamic critical exponents of cluster algorithms
are well-studied by using the autocorrelation times of spin systems in
thermal equilibrium~\cite{SwendsenWang,Wolff,Ito1990,Wolff1989,Heerman1990,Baillie1991,Tamayo1990}.
The autocorrelation time ($\tau_W$) for the Wolff's algorithm can be
obtained by the relation
\begin{equation}
	\tau_W \;=\; {\tau_W}^{\prime} \frac{<C>}{L^d} \;  .
\end{equation}
Considering the finite size scaling behavior of the average cluster
size, $\frac{<C>}{L^d} \sim L^{2(Y_H - d)}$, the dynamic critical
exponent ($z$) can be calculated. In these studies, the dynamic
critical exponent is observed to be much less than that of local
algorithms. For measurements done in thermal equilibrium, since the
correlation length is as large as the lattice size, and the
fluctuations are at their maxima, finite size effects create
difficulties in extracting the dynamic critical exponent. Hence for
the measurement of the dynamic critical exponent, very good statistics
and very large lattices are essential in order to obtain accurate
results.  For the $2$-dimensional Ising model, Wolff showed that $z\sim
0.25$~\cite{Wolff1989}.  More recently, it is suggested that data is
consistent with a logarithmic divergence~\cite{Heerman1990}, but it is
very difficult to distinguish between a logarithm and a small
power~\cite{Baillie1991}.  For the $3$-dimensional case, Tamayo et
al.~\cite{Tamayo1990} obtained the dynamic critical exponent as $z\sim
0.44(10)$.  Wolff (by using energy autocorrelations) have calculated a
smaller value of $z=0.28(2)$~\cite{Wolff1989}.  In $4$-dimensions,
Tamayo et al.~\cite{Tamayo1990} obtained $z$ with a vanishing
value. This result is also consistent with the mean-field solution for
the Ising model in four and higher dimensions.
Recently various cluster algorithms have been tested~\cite{Wang2002}
and compared to the original versions of Swendsen-Wang and Wolff's
algorithms. In this study critical exponents of  $2$- and $3$- dimensional
Ising models have been observed to be small and the $z$ values are 
consistent with the above mentioned values.   

$\;$

In the quenching process of the spin systems, very small clusters
start to grow until the cluster size reaches the cluster size of the
quenching temperature. Hence for the Wolff's algorithm one expects a
very low efficiency at the beginning of the simulation and, as the
iterations and the cluster sizes are increased, the efficiency of the
algorithm grows.  Since the efficiency of the Wolff's algorithm is
directly related to the size of the updated clusters, one expects that
continuously increasing efficiency can be seen in dynamic processes.
Considering the number of updated spins at each iteration, one can
estimate the dynamic critical exponent which is characteristic of the
time scale of the system. The expectation that the dynamic critical
exponent is the same for both the early stages of the simulation and
in thermal equilibrium (in analogy with the earlier
work~\cite{Zheng98,Jaster99,Zheng00,Luo97,Ying2001,Dilaver2003}) is the
main motivation of this work. In this work we aimed to discuss dynamic
scaling and to obtain the dynamic critical exponent of the Wolff's
algorithm by using dynamic finite size scaling relation
(Eq.(\ref{Magnetization})). We plan to use the known critical
exponents. Using these critical exponents and observing a good scaling 
behavior requires no or very little size corrections.

\section{The Model and the Simulations}

In this work we have employed $2$- $3$- and $4$-dimensional Ising models
described by the Hamiltonian
\begin{equation}
	\label{hamiltonian}
	{-\beta}H = K{\sum_{<ij>}}{S_{i}}{S_{j}} \;.
\end{equation}
Here, $\beta=1/kT$ and $K=J/kT$, where $k$ is the Boltzmann constant,
$T$ is the temperature and $J$ is the magnetic interaction between the
spins. In the Ising model the spin variables take the values
${S_{i}}={\pm 1}$. 

$\;$

In order to discuss time evolution of the  Wolff's algorithm, we have
selected a range of thermodynamic quantities. Equation (\ref{Magnetization})
implies that the magnetization ($<S>$) and its higher moments are good
candidates for observing dynamic finite size scaling behavior. For
this reason we have considered $<S>$, $<S^2>$ and $<S^4>$ where, the
$n^{th}$ moment of the magnetization is given by
\begin{equation}
	\label{MomentsOfSpin}
	<S^n> = {1 \over L^d} < (\sum_i S_i)^n > \;.
\end{equation}
The efficiency of the Wolff's algorithm is related to the
average cluster size ($<C>$), with
\begin{equation}
	\label{AverageClusterSize}
	< C >\; =\;{1\over{N_{c}}}{\sum_{i}^{{N_{c}}}}{1\over{L^d}}{(C_{i})},
\end{equation}
and it has been considered in our calculations.  For
finite size lattices, during the quenching process, very small
clusters grow to the lattice size as the number of iterations
increases.  The rate of the cluster growth depends on the algorithm as
well as the quenching temperature.  More efficient the algorithm,
faster the system reaches the equilibrium values.  In a
single-cluster Wolff's algorithm, 
one must consider the average number of updated spins, or the
percentage of them, considering total volume. Since the
number of updated spins increases at each iteration (as the system approaches the
quenching temperature), repetitive calculations yield an average
cluster size per iteration. Consequently in calculating the
autocorrelation time, one considers the number of updated spins per
iteration.  In our calculations, consistent with
Eq.(\ref{Magnetization}) both moments of spin
(Eq.(\ref{MomentsOfSpin})) and the average cluster size
(Eq.(\ref{AverageClusterSize})) are calculated as functions of
iteration number.
In the time-dependent forms of the above quantities,
Eq.(\ref{MomentsOfSpin}) and Eq.(\ref{AverageClusterSize}) take the form
\begin{equation}
	\label{MomentsOfSpinOft}
	<S^n>(t) = {1 \over L^d} < (\sum_i S_i(t))^n >
\end{equation}
and
\begin{equation}
	\label{AverageClusterSizeoft}
	< C >(t)\; =\;{1\over{N_{c}}}{\sum_{i}^{{N_{c}}}}{1\over{L^d}}{(C_{i}(t))}.
\end{equation}

As it is mentioned above, in this algorithm efficiency varies and effective time changes
(during the quenching process) are related to the quantities
$<C>(t)$ and $<S^2>(t)$.

\vskip\baselineskip

All of the above quantities have their own anomalous dimensions and
using such quantities (in order to obtain dynamic critical exponent) one may
expect some ambiguities due to correction to scaling. Since our
calculations are done in the early stages of the simulations, the
correlation length is expected to be less than the lattice size; hence
use of infinite lattice critical exponents in Eq.(\ref{Magnetization})
can be sufficient to explain the critical behavior of the system. For
this reason, critical exponents are taken as the Onsager solution for
the $2$-dimensional Ising model. For the $3$-dimensional case, the
critical exponent values are taken from the
literature~\cite{Blote95,Talapov96}. The $4$-dimensional case is the
critical dimension for the Ising model, and above $4$-dimension the
critical exponents are the mean-field critical exponents. In
the $4$-dimensional case, the model possesses the mean-field critical
exponents with logarithmic corrections. In our calculations no
corrections are assumed to be necessary, since during the early stages
of the quenching process the correlation length remains smaller than
the lattice size.

$\;$

In this work we have studied $2$-, $3$- and $4$-dimensional Ising
models evolving in time by using Wolff's algorithm. We have prepared
lattices with vanishing magnetization and total random initial
configurations are quenched at the corresponding infinite lattice
critical temperature. We have used the lattices $L=256$, $384$, $512$, $640$,
$L=32$, $48$, $64$, $80$ and $L=16$, $20$, $24$, for $2$-, $3$- and $4$-dimensional
Ising models, respectively. For each dimension and size, iterations are
continued until the observed quantity reaches a plateau which
finalizes the evolution during the quenching process. In order to
observe the plateau, the number of iterations has been chosen depending
on the volume of the system.  For each lattice size, initial
configurations are created depending on the lattice size, but on
average sixteen bins of eight to thirty thousand runs, sixteen bins of
four to thirty five thousand runs and sixteen bins of five to
twenty five thousand runs have been performed for $2$-, $3$-, and $4$-
dimensional Ising models for varying lattice sizes, respectively. Errors are
calculated from the average values for each iteration obtained in
different bins. Both the average cluster size and susceptibility have
the same anomalous dimension, hence in obtaining the autocorrelation time ($\tau$) from the
observed behavior of the dynamic variable, one can replace $<S^2>$ by
$<C>$. In our simulation data both quantities have been used in order
to scale time variable for the scaling of average cluster size ($<C>$)
and moments of the magnetization ($<S^n>$).

$\;$

In Figure 1 we have presented the magnetization data $(<S>(t))$ before
and after the dynamic finite size scaling for $2$-dimensional Ising
model for the lattice sizes considered. Figure 1(a) shows the time
evolution of $<S>(t)$ during the relaxation of the system until a
plateau is reached. It is seen from this figure that the time to reach
the plateau is proportional to the linear size $(L)$ of the system. As
it is seen from Eq.(\ref{Magnetization}), in the dynamic finite size
scaling, $<S>(t)$ scales with factor $L^{(Y_H-d)}$ and, for the
algorithms in which all spins are checked for updating the Monte Carlo
time, $t$ scales as $t/{L^z}$. For Wolff's algorithm, one obtains
crossing magnetization curves but they do not scale as expected.
Since one cluster is updated at each iteration, there is a need to use
the average number of updated spins at each iteration, which is a
factor related to $<C>(t)$ or $<S^2>(t)$. If the time is not scaled by
the average cluster size, using only $L^z$ as a factor shifts the
curves towards each other and curves crosses at some point, but
scaling can not be observed. Figure 1(b) and (c) show the scaling of
Monte Carlo time axis by using $<C>(t)$ and $<S^2>(t)$ as the factor
of scaling, respectively. The fluctuations in Figure 1(c) are
smaller, compared to the fluctuations in Figure 1(b). In the following
figures we have presented the scaling using $<S^2>(t)$ as the factor
for time scaling. Magnetization, second and fourth moments of the
magnetization, as well as the average cluster size show the same
scaling behavior in all dimensions; but in order to avoid repetitive
figures, we have chosen $<S>(t)$ as an example. Only for the
$2$-dimensional case we have presented $<S^2>(t)$ in Figure 2 for
comparison.

$\;$

Figure 2 shows the dynamic quantity $<S^2>(t)$ for the $2$-dimensional
Ising model for the same lattice sizes given in Figure 1. Figure 2(a)
shows the simulation data and Figure 2(b) shows the scaling by use of
$<S^2>(t)$ as the factor in time scaling. As observed for $<S>(t)$,
scaling using $<C>(t)$ and $<S^2>(t)$ as the factor in time scaling
results in the same value of the dynamic critical exponent $z$.  As it
is mentioned above the average cluster size and the moments of the
magnetization are calulated as averages at each iteration and these
quantities are time-dependent during the relaxation process.  In
Figure 2 same factor $<S^2>$ appears on both the vertical and
horizontal axes. Since $t$ increases linearly, the curve in Figure 2
is linear. This also shows that the scaling is very good. Figures 3
and 4 show the simulation data and the dynamic scaling for $<S>(t)$
for $3$- and $4$- dimensional Ising models, respectively. In both
figures a) shows the time evolution of $<S>(t)$ and b) shows $<S>(t)$
after dynamic scaling.

$\;$

The values of the dynamic critical exponent $z$ for $2$-, $3$- and $4$-
dimensional Ising models are given in Table 1. In this table $z^{\prime}$ is
the measured dynamic critical exponent as an average of the exponents
calculated for $<S>(t)$, $<S^2>(t)$, $<S^4>(t)$ and $<C>(t)$.
The dynamic critical exponent $z$ is calculated using
Eq.(\ref{Magnetization}) by considering the number of spins updated at
each iteration of the Monte Carlo simulation and by using the scaling
relation for time ($\tau \sim L^z$). It is a simple relation in the form  

 \begin{equation}
z=z^{\prime}-(2Y_H-d)  .
\end{equation}

In this calculation, $Y_H$ is taken as $Y_H={15\over 8}$ (Onsager
solution), $Y_H=2.4808$~\cite{Blote95,Talapov96}, $Y_H=3$
(mean-field solution) for the $2$-, $3$- and $4$- dimensional
models respectively.

$\;$

The errors in the values of $z^{\prime}$ are obtained from the largest
fluctuations in the simulation data for $<S>(t)$, $<S^2>(t)$, $<S^4>(t)$ and
$<C>(t)$. By considering the values of the dynamic critical exponents as
the averages obtained using scaling of these four quantities and the
small errors in $z$ values, one can see from Table 1 that the
scaling behavior is quite similar for these four quantities.

\section{Conclusion}

In this work we have observed the dynamic finite size scaling for
$2$-, $3$- and $4$- dimensional Ising models. In previous
works~\cite{Zheng98,Zheng00,Jaster99}, the dynamic finite size
scaling of the Ising model has been tested by using Metropolis and the
heat bath algorithms. Both algorithms are local algorithms, hence the
observed similar behavior is expected. In our work we have shown for
the first time that dynamic finite size scaling, exhibiting a
universal behavior, is valid for use with cluster algorithms, and that
both local and global algorithms show dynamic finite size scaling with
their own dynamic critical exponents. In this work we have calculated
the dynamic critical exponent of Wolff's cluster algorithm for $2$-,
$3$- and $4$- dimensional Ising models using dynamic scaling. In the
literature, for all three dimensions, small dynamic critical exponents are
obtained~\cite{SwendsenWang,Wolff,Ito1990,Wolff1989,Heerman1990,Baillie1991,Tamayo1990}, 
but further studies of the data suggest that for
all three dimensions the dynamic critical exponent of the Ising model
can be considered as zero.  The measurement of the dynamic critical
exponent in thermal equilibrium is extremely difficult since the
correlation length around the phase transition point is as large as
the size of the lattice. In dynamic finite size scaling, since the
correlation length remains smaller than the lattice size, it is
expected that statistically independent configurations lead to better
statistics since there is no finite size effects. In our calculations,
we have observed that our data is consistent with vanishing dynamic
critical exponent. Despite the fact that obtaining good statistics is
extremely time consuming for large lattices, data is free of errors
due to finite size effects. One can see from the results of dynamic
scaling that scaling is very good and the errors are very small, hence
this method is a good candidate to calculate the dynamic critical
exponent for any spin model and for any algorithm.

\section*{Acknowledgments}

Y. G. and S. G. thank the Physics Department at Pisa University,
where some of the analysis have been done, for their hospitality. 
Authors greatfully acknowledge Hacettepe University Research Fund
\mbox{(Project no : 01 01 602 019)} and Hewlett-Packard's Philanthropy Programme.
We thank Dale Ross for carefully reading the manuscript.

\pagebreak

\pagebreak

\section*{Table Captions}

Table 1.  The values of measured ($z^{\prime}$) and calculated ($z$) 
dynamic exponents for $2$-, $3$- and $4$-dimensional Ising models.

\section*{Figure Captions}

Figure 1 a) Magnetization data $<S>(t)$ for $2$-dimensional Ising
Model for linear lattice sizes $L=256$, $384$, $512$, $640$ as a function of
simulation time $t$, b) scaling of $<S>(t)$ data given in a) using
$<C>(t)$ as the factor in time scaling, c) scaling of $<S>(t)$ data
given in a) using $<S^2>(t)$ as the factor in time scaling.

$\;$

Figure 2 a) Simulation data for $<S^2>(t)$ as a function of simulation
time $t$ for $2$-dimensional Ising model for linear lattice sizes
$L=256$, $384$, $512$, $640$, b) scaling of $<S^2>(t)$ data given in a) using
$<S^2>(t)$ as a factor in time scaling.

$\;$ 

Figure 3. Simulation data for $<S>(t)$ as a function of simulation
time $t$ for $3$-dimensional Ising model for linear lattice sizes
$L=32$, $48$, $64$, $80$, b) scaling of $<S>(t)$ data given in a) using
$<S^2>(t)$ as the factor in time scaling.

$\;$

Figure 4. Simulation data for $<S>(t)$ as a function of simulation
time $t$ for $4$-dimensional Ising model for linear lattice sizes
$L=16$, $20$, $24$, b) scaling of $<S>(t)$ data given in a) using $<S^2>(t)$
as the factor in time scaling.

$\;$

\pagebreak

\begin{center}
\begin{tabular}{|p{2cm}|p{3cm}|p{5cm}|}
\hline
$~~~~~~d$   &  ~~~~~~~~~~$z^{\prime}$  &  ~~~$z = z^{\prime} - (2 Y_H - d)$\\
\hline
$~~~~~~2$     &  ~~~~$ 1.73 \pm 0.05$      & ~~~$ ~~0.02 \pm 0.05 $\\
\hline
$~~~~~~3$   & ~~~~$1.94 \pm 0.09  $       & ~~~$ ~~0.02 \pm 0.09 $ \\
\hline
$~~~~~~4$    & ~~~~$ 2.13 \pm 0.19 $       & ~~~$ -0.13 \pm 0.19$   \\
\hline
\end{tabular}\\
\vskip 0.5cm
\centerline {Table 1.}  
\end{center}

\begin{figure}
\centering
\subfigure[]{\includegraphics[angle=0,height=4.3truecm]{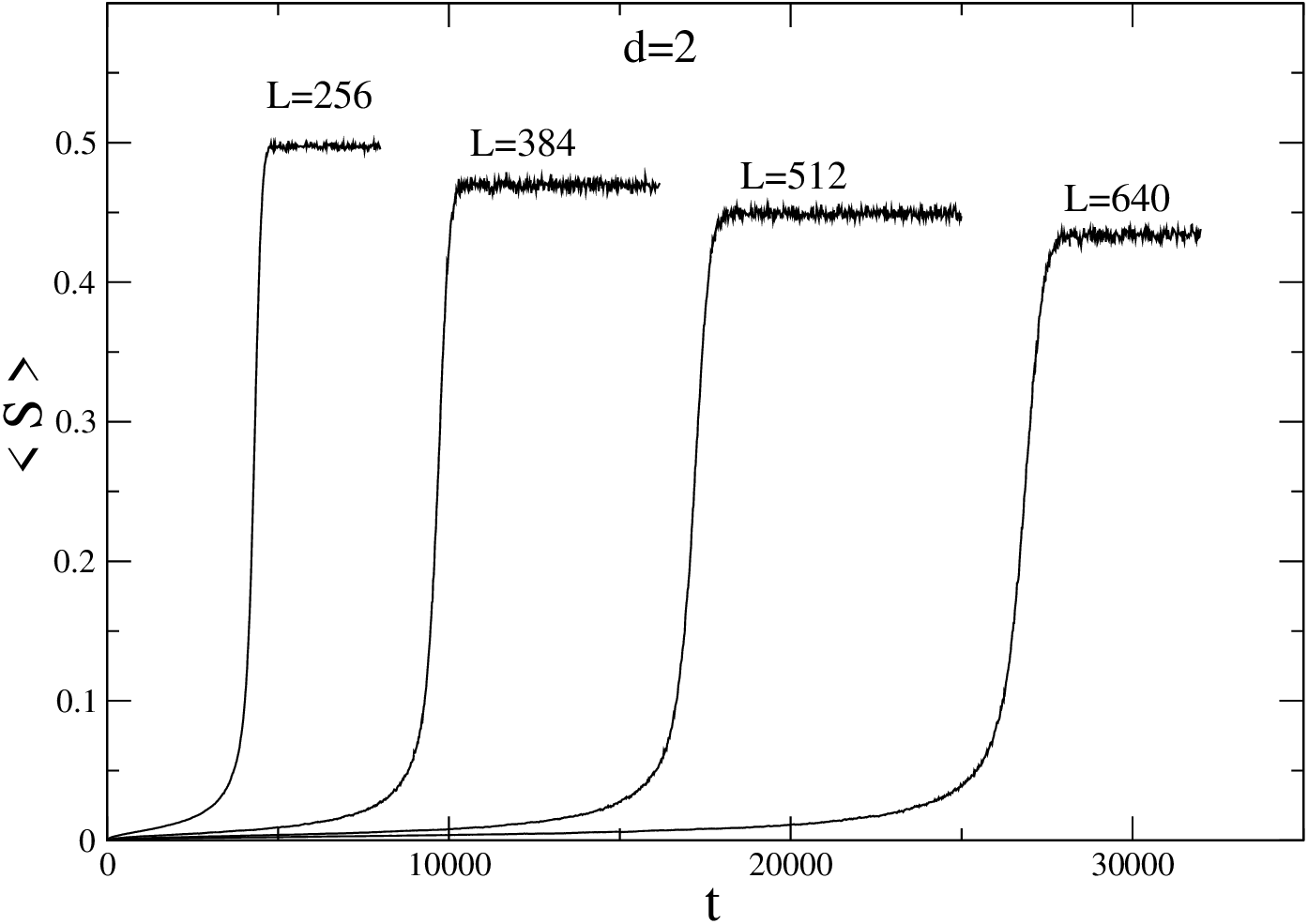}}\\
\subfigure[]{\includegraphics[angle=0,height=4.3truecm]{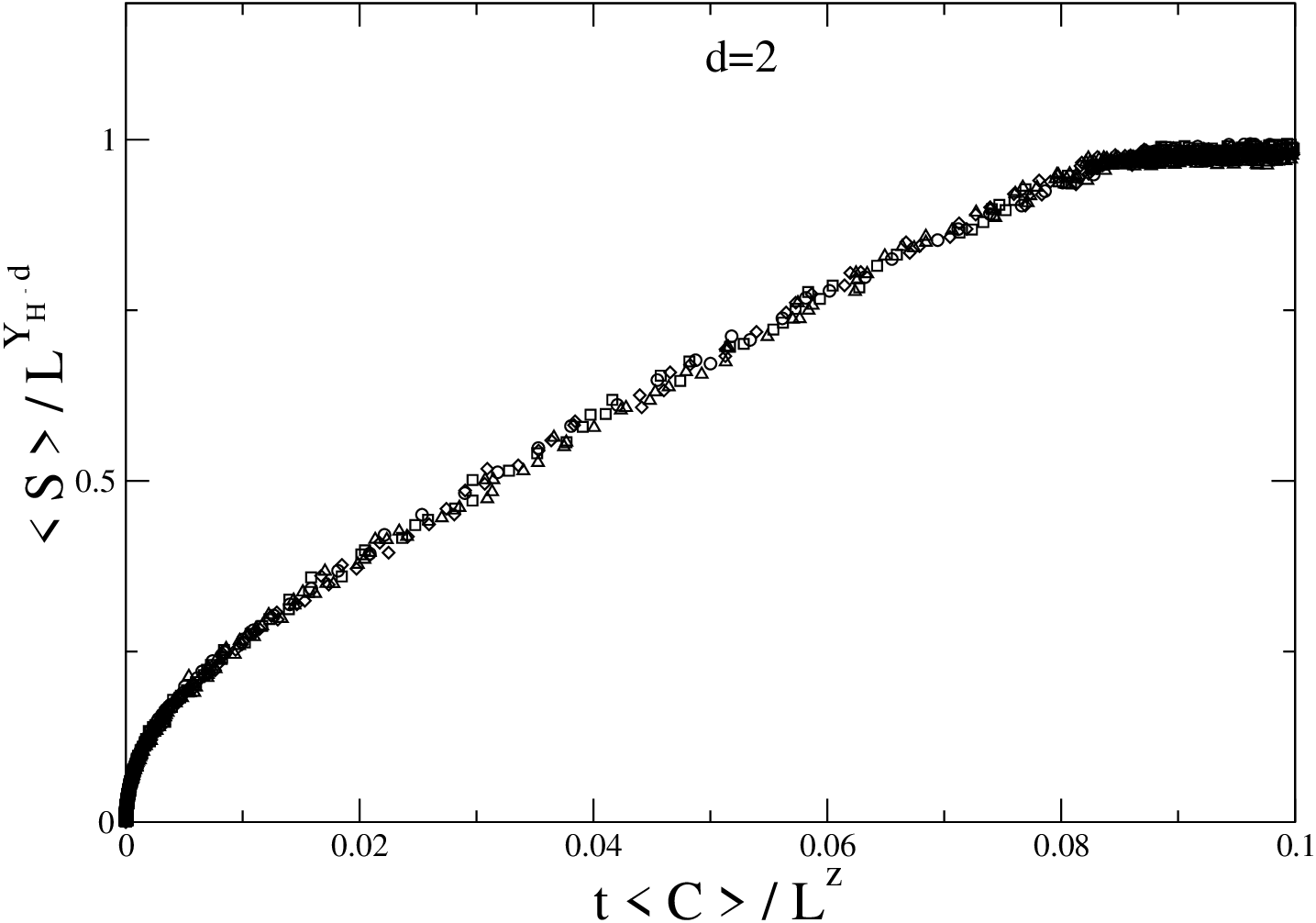}}\\
\subfigure[]{\includegraphics[angle=0,height=4.3truecm]{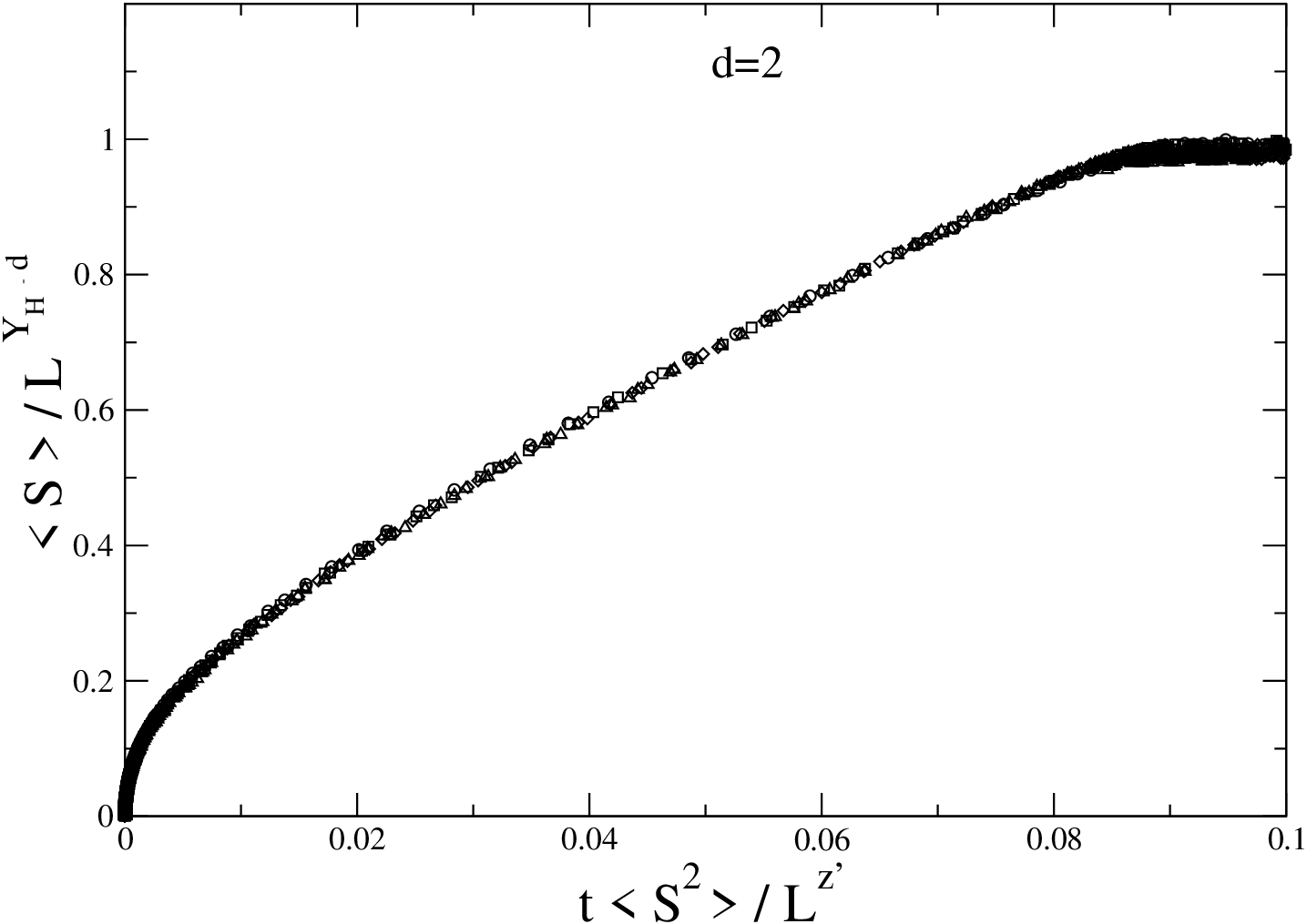}}
\caption{}
\label{fig1}
\end{figure}

\begin{figure}
\centering
\subfigure[]{\includegraphics[angle=0,width=6truecm]{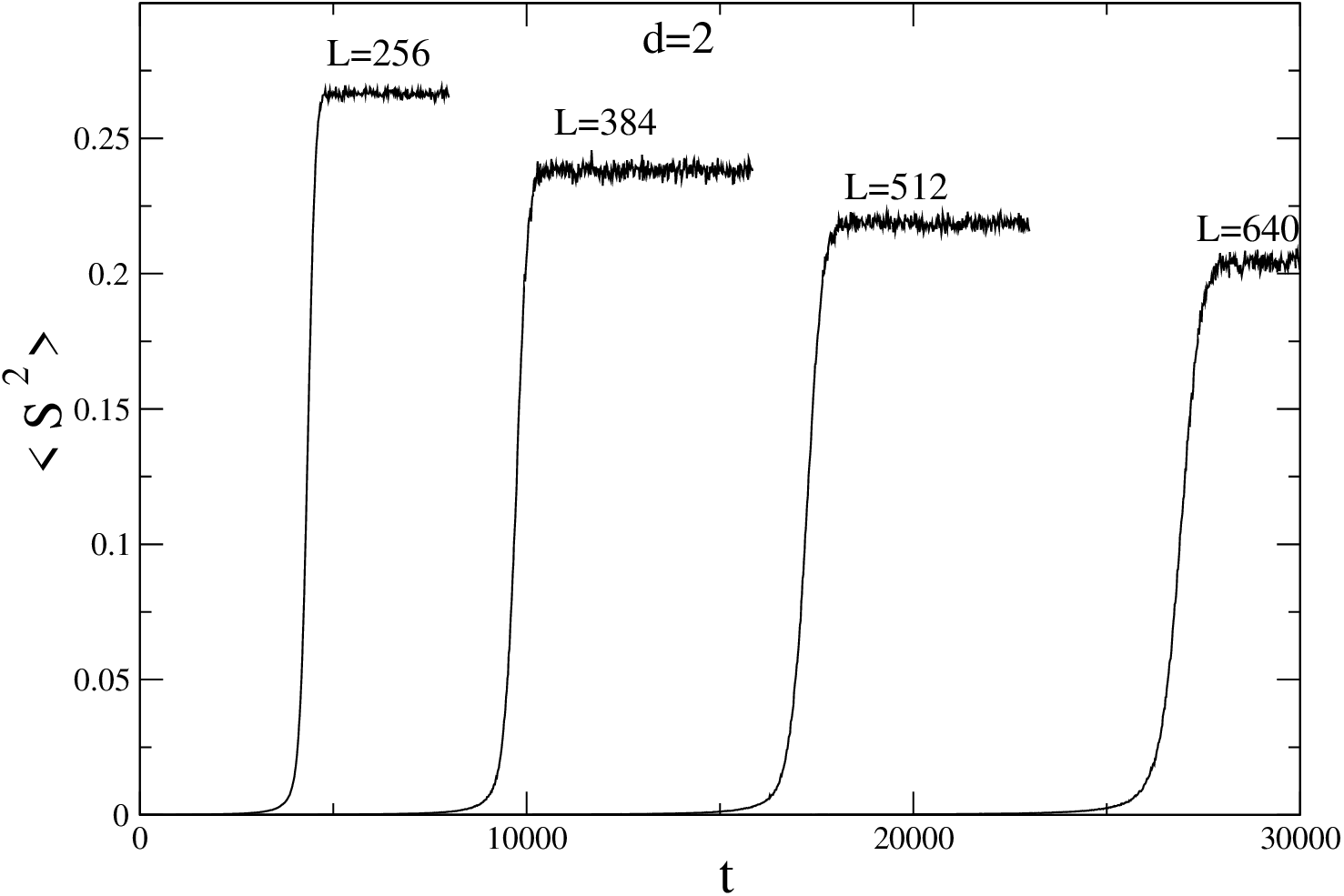}}
\subfigure[]{\includegraphics[angle=0,width=6truecm]{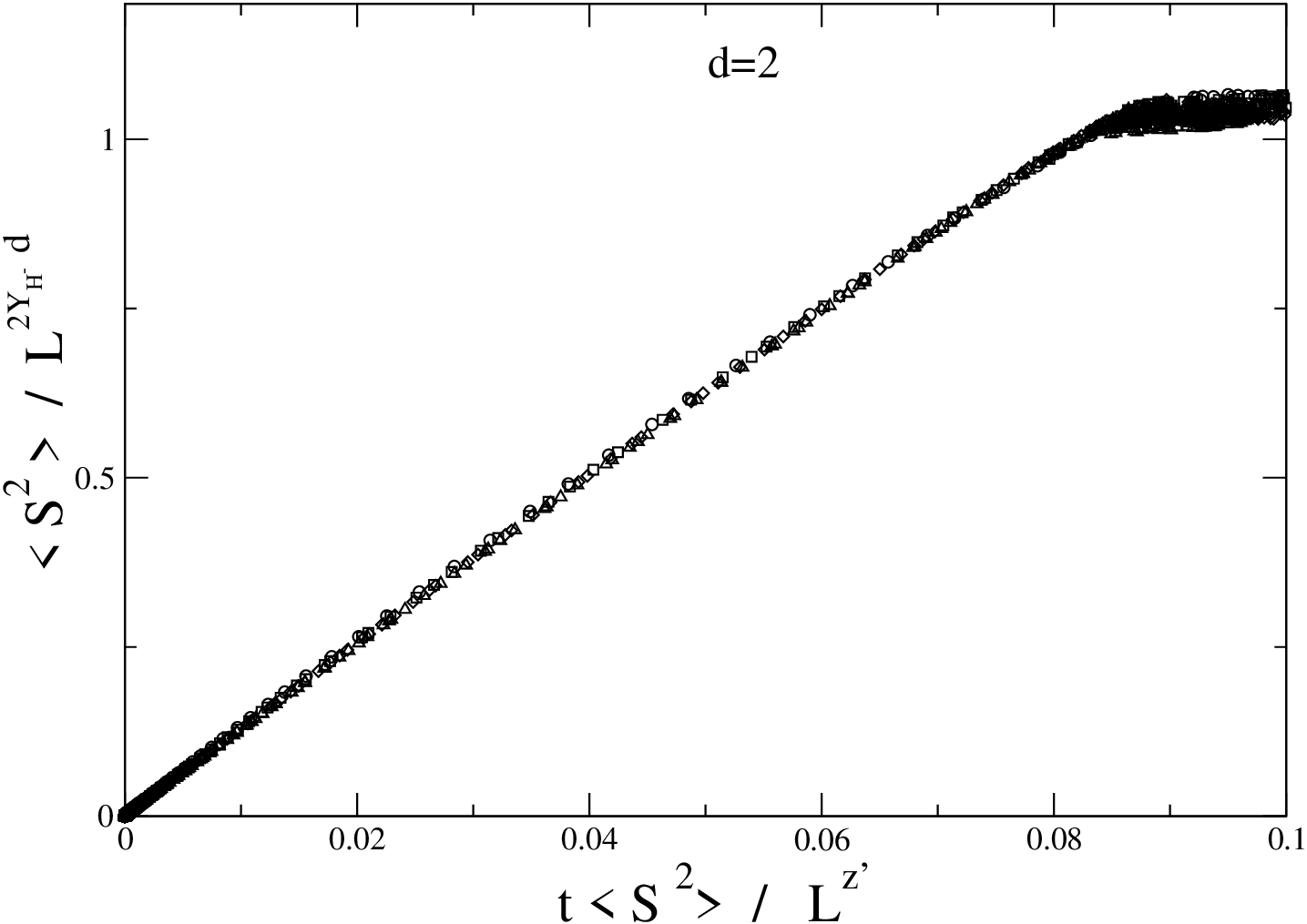}}
\caption{}
\label{fig2}
\end{figure}

\begin{figure}
\centering
\subfigure[]{\includegraphics[angle=0,width=6truecm]{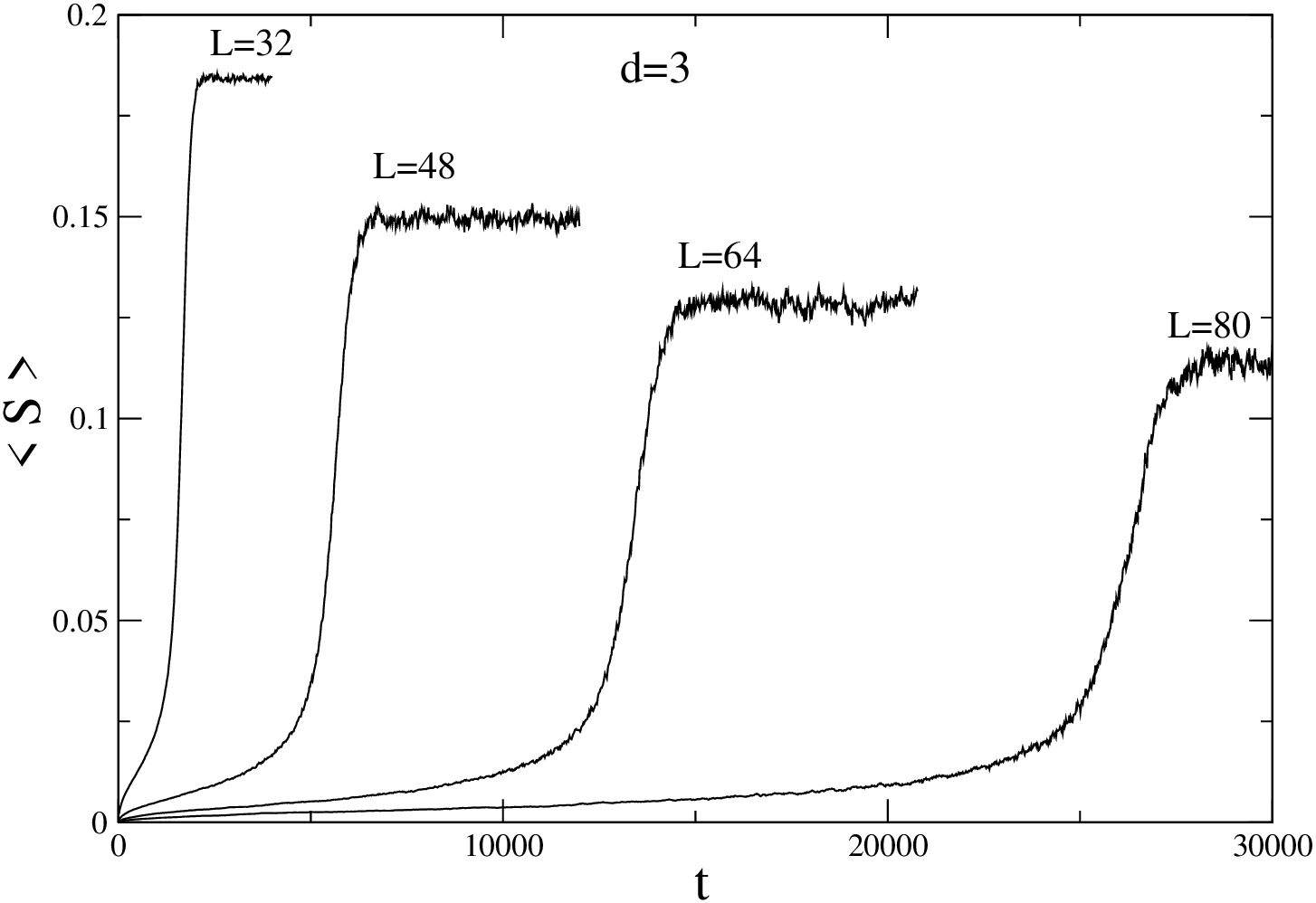}}
\subfigure[]{\includegraphics[angle=0,width=6truecm]{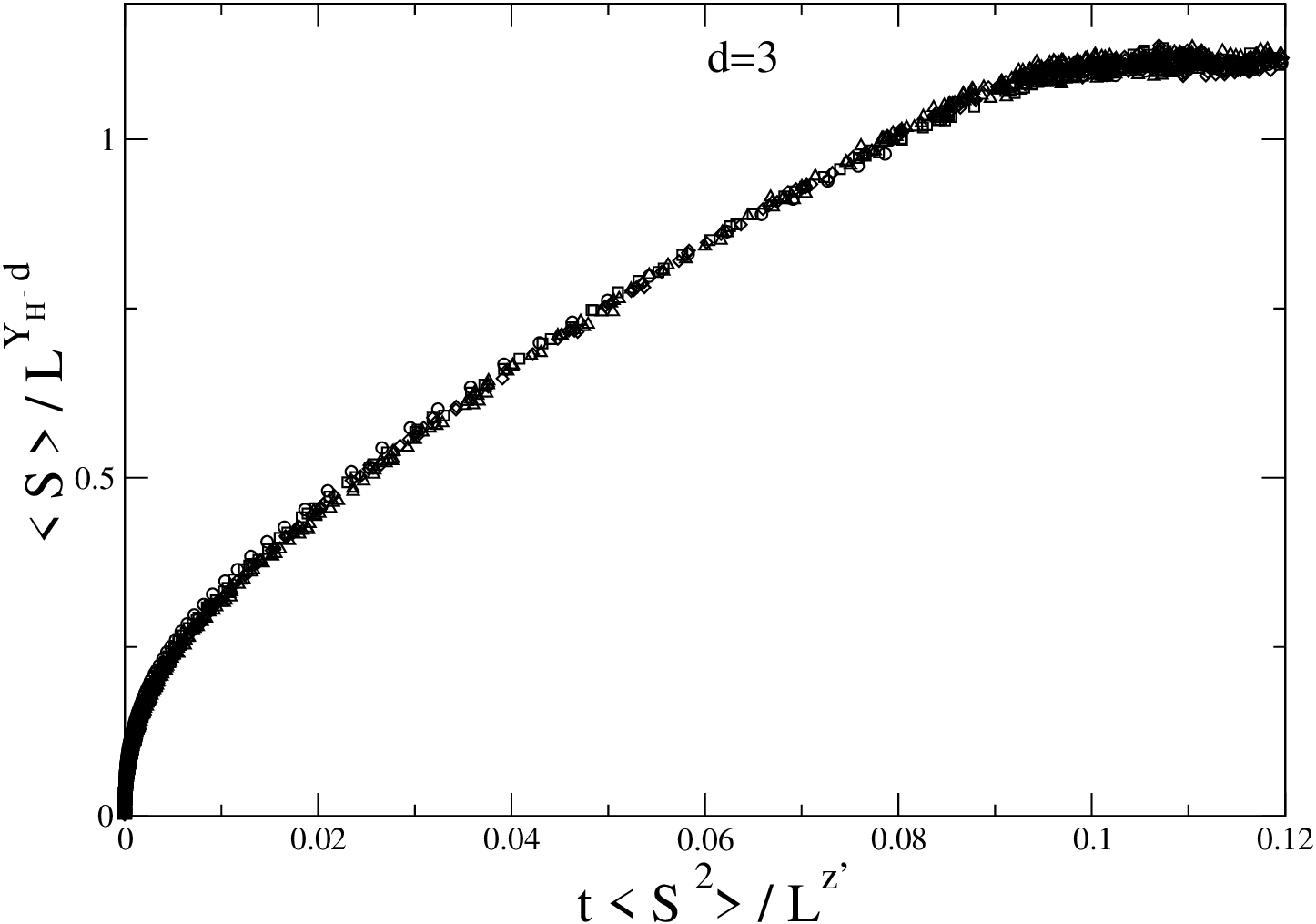}}
\caption{}
\label{fig3}
\end{figure}

\begin{figure}
\centering
\subfigure[]{\includegraphics[angle=0,width=6truecm]{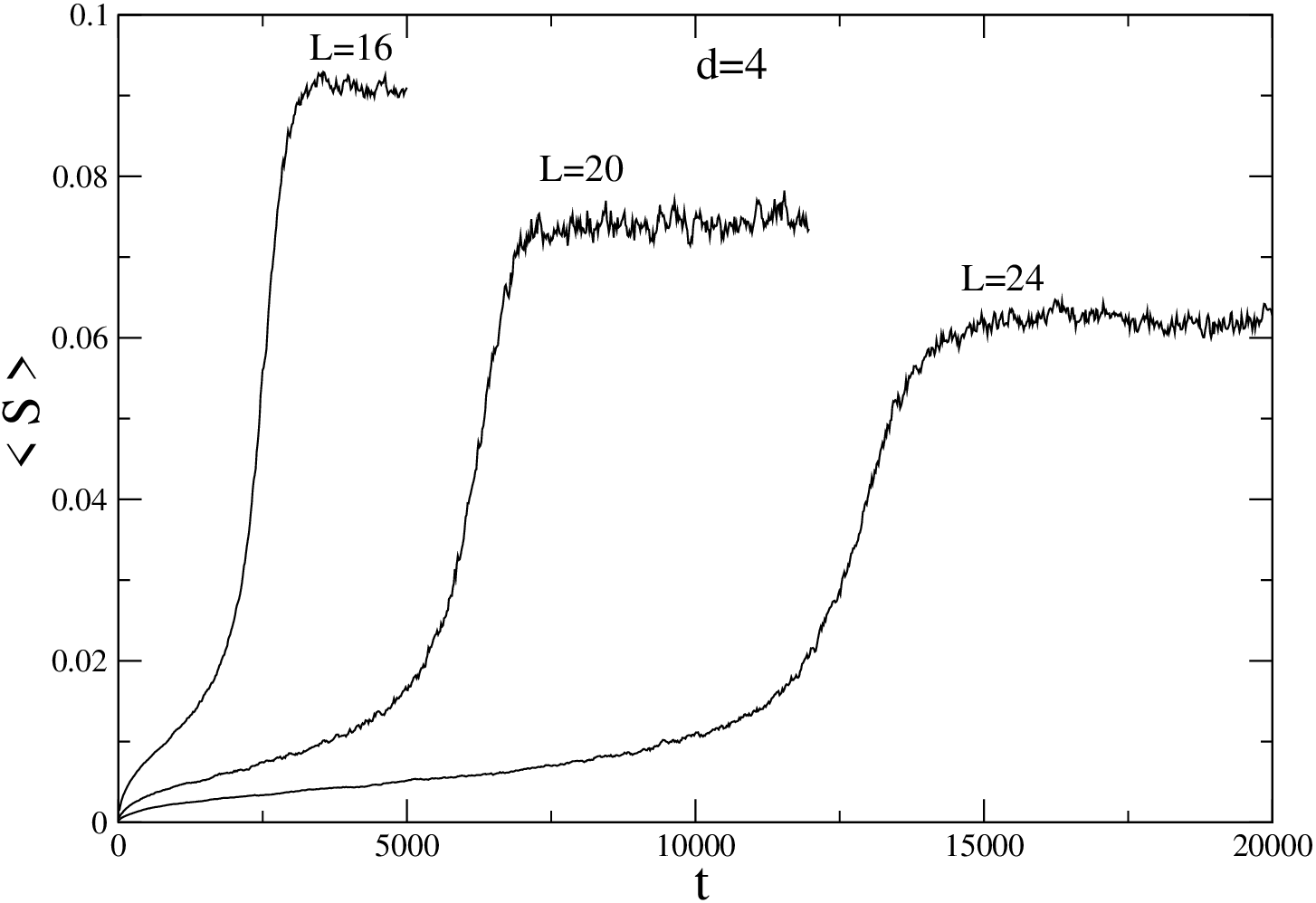}}
\subfigure[]{\includegraphics[angle=0,width=6truecm]{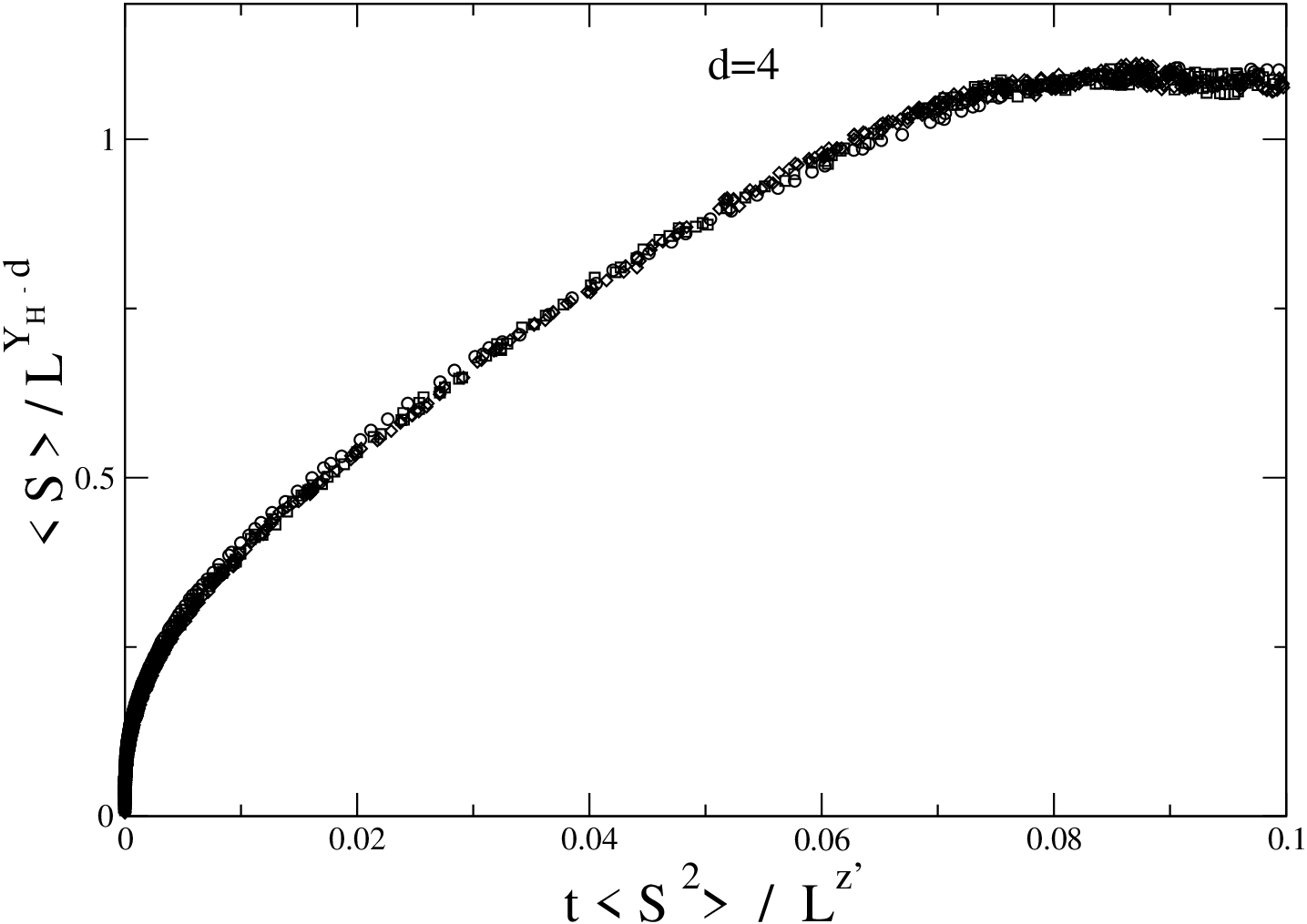}}
\caption{}
\label{fig4}
\end{figure}

\end{document}